# Structure, corrosion resistance and nanomechanical properties of CoCrFeNiX (X=Nb,Mo,B,Si) high entropy alloys

R. Babilas[1,*], J. Bicz[1], A. Radoń[2], M. Kądziołka-Gaweł[3], D. Łukowiec[4], K. Matus[4], E. Wyszkowska[5],

Ł. Kurpaska[5], D. Rudomilova[6], K. Młynarek-Żak[7]

[1] Department of Engineering Materials and Biomaterials, Silesian University of Technology, Konarskiego 18a St., Gliwice 44-100, Poland
[2] Łukasiewicz Research Network – Institute of Non-Ferrous Metals, Sowińskiego 5 St., 44-100 Gliwice, Poland
[3] Institute of Physics, University of Silesia in Katowice, 75 Pułku Piechoty 1, 41-500 Chorzów, Poland
[4] Materials Research Laboratory, Silesian University of Technology, Konarskiego 18a St., 44-100 Gliwice, Poland
[5] National Centre for Nuclear Research, NOMATEN CoE MAB+, Andrzeja Soltana 7 St., 05-400 Otwock-Świerk, Poland
[6] Technopark Kralupy, University of Chemistry and Technology Prague, 166 28 Prague, Czech Republic
[7] Department of Engineering Processes Automation and Integrated Manufacturing Systems, Silesian University of Technology, Konarskiego 18a St., Gliwice 44-100, Poland

**Corresponding author:** rafal.babilas@polsl.pl

**Abstract**

In this work, the four high entropy CoCrFeNiX alloys (X=Mo,Nb,B,Si) were prepared by induction melting to comparatively analyse their structure, nanomechanical properties, and corrosion resistance in the chloride ion environment. The CoCrFeNiNb and CoCrFeNiMo alloys are composed of FCC solid solution and intermetallic phases $(TM)_2Nb$ and Cr-Mo-TM. In the case of the CoCrFeNiB alloy, a complex phase structure was revealed, consisting of FCC solid solution and three types of borides. In turn, the addition of Si substantially altered the phase composition of the CoCrFeNi alloy, resulting in the formation of two intermetallic phases. The corrosion behaviour of the alloys was studied in 3.5 and 5% NaCl solutions. The highest corrosion resistance in both solutions used characterize the CoCrFeNiSi alloy, showing the lowest corrosion current density and the most positive corrosion potential values. For measurements in 5% NaCl solution, $i_{corr}$ and $E_{corr}$ were equal to 0.24 µA/cm$^2$ and -0.136 V. Currently, the least favourable corrosion parameters were recorded for the CoCrFeNiMo alloy. The results of EIS measurements confirmed the high protective abilities of passive film formed on the CoCrFeNiSi alloy surface. The highest strength properties were shown by the alloys with the addition of metalloids. For the CoCrFeNiSi alloy, the highest nanohardness value was obtained (above 15 GPa), while the CoCrFeNiB showed the highest Young modulus (above 275 GPa).

**Keywords**

## 1. Introduction

The high-entropy alloys (HEAs) represent a new research frontier in metallurgy. Differently from the traditional approach to alloy development, their design requires the use of five or more elements, in concentrations between 5 and 35 at.%. As a result, these alloys exhibit significantly higher configurational entropy, which contributes to enhanced mutual solid solubility of the constituent elements, thus inhibiting the formation of intermetallic phases. This characteristic, called the 'high entropy effect', results in a significant simplification of its microstructure, expanding the possibilities of synthesis, processing and application [1-3]. HEAs exhibit unique properties, such as high hardness [4], superior high-temperature strength [5], excellent corrosion resistance [6] or irradiation resistance [7]. Furthermore, HEAs can exhibit catalytic properties, creating opportunities for their use in wastewater treatment [8]. For example, in the work [4], a possibility of the AlCoCrFeNiTi and AlCoCrFeNiTiSi alloys application as catalysts in the Fenton reaction, used for degradation of the chemical dye – Rhodamine B, was investigated. The results indicated that both alloys examined show high catalytic activity, allowing one to obtain, respectively, 92 and 95.5% decolourisation rate after 60 minutes of process conducted under optimum reaction conditions – as compared to 52% obtained for the reference reaction.

The CoCrFeNi alloy with the face-centred cubic structure exhibits advantageous properties, such as excellent fracture toughness at a wide range of temperatures and good corrosion resistance. However, its strength is insufficient for many structural applications [9,10]. Various strengthening mechanisms are being employed to improve the strength properties of FCC-type HEAs, including solid solution strengthening, grain boundary strengthening and precipitation strengthening [10,11].

The method often applied is to introduce refractory elements, such as Mo, Nb or W [9,11-18]. Their addition, as a result of larger atomic radii, causes an increase in lattice distortion, resulting in the strengthening of the solid solution, and in larger quantities, contributes to precipitation of hard intermetallic phases, such as the Laves phase for $CoCrFeNiNb_x$ [15, 16] or $\mu$ phase for $CoCrFeNiW_x$ [19]. The synergistic effect of solid solution and precipitation strengthening leads to a significant improvement in mechanical properties [14,17]. For example, the introduction of Mo into the equimolar CoCrFeNi alloy results in the evolution of the microstructure, causing the formation of an intermetallic $\sigma$ phase and, subsequently, with further increase in the Mo content, the appearance of the intermetallic $\mu$ phase. In effect, its compressive strength increased from 871 to 1441 MPa, however, at the expense of ductility – fracture strain decreased from 75 to 21% [14]. In turn, in the case of the $CoCrFeNiNb_{0.45}$ alloy, an exceptional compressive strength was obtained – exceeding 2550 MPa, with simultaneously retaining 27.9% of the fracture strain, which is attributed to its nanolamellar eutectic microstructure [18].

Another promising strategy to improve the mechanical properties is an introduction of metalloid elements, such as boron or silicon [10,20,21]. The authors of the work [20] reported that microalloying with boron allows an effective increase in the tensile strength of the CoCrFeMnNi alloy after recrystallisation annealing, while simultaneously retaining its high ductility. Boron was found to segregate on the grain boundaries, affecting its energy, structure, and grain size. Boron doping results in increase of the activation barriers for grain coarsening, in effect stabilizing the grain size during recrystallization. Furthermore, also from the solute, boron improves the cohesion of the grain boundaries under loading conditions [20,21]. As a result, the addition of boron contributes to grain boundary strengthening effect, allowing one to enhance the ultimate tensile strength by 40% for the alloy doped with 30 ppm of boron. As the solid solubility of boron in CoCrFeMnNi is quite limited, the addition of this element at a higher concentration causes the formation of borides in the grain boundaries, negatively affects the mechanical properties [20]. In turn, Zhang et al. [22] reported that the addition of B to the CoCrFeNi alloy changes its plastic deformation mechanisms, which can be associated with its influence on stacking fault energy (SFE). The minor addition of boron contributes to the decrease in the SFE, altering the main deformation mechanism from dislocation slip to mechanical twinning. An increase in boron content is observed to reverse this effect, which can be attributed to the formation of borides, involving the chromium of the solid solution phase – leading to an increase in SFE [22]. The appearance of deformation twins introduces additional barriers to dislocation movement, reducing their mean free path, resulting in a strengthening effect [10,23].

In the work [10] the effect of the addition of silicon on the microstructure and properties of the CoCrFeNi alloy subjected to thermomechanical processing was studied. It was found that the single phase face-centred cubic structure of the alloy can be retained for its concentrations up to 10 at.%. As a result of its smaller atomic radius – compared to other constituent elements - the introduction of silicon to the CoCrFeNi alloy increases lattice distortion, contributing to the solid solution strengthening effect. Furthermore, it was reported that silicon addition affects plastic deformation mechanisms, which is connected with its effect on the stacking fault energy. The decrease in SFE improved the role of mechanical twinning, contributing to an advantageous combination of strength and ductility [10]. A similar effect was observed for the CoCrFeMnNi alloy [24], which allowed the concurrent enhancement of its strength and ductility. It was shown that alloying with 10 at.% of silicon contributed to increase its ultimate tensile strength by 45%, with simultaneous improvement in ductility. In the case of the CoCrFeNi alloy, it was further found that for 10 at.% silicon addition, the deformation mechanisms also involve the transformation induced plasticity effect (TRIP), associated with the transformation of the cubic to hexagonal close-packed phase [10].

The possibility of forming a stable and uniform passivation film is the key factor influencing the corrosion resistance of alloys [25]. In the case of the CoCrFeNi alloy, the presence of chromium and nickel plays a decisive role, contributing to its advantageous properties – superior compared to conventional AlSl 304 stainless steel [26]. The significant factor that influences the corrosion behaviour of HEAs is the high entropy effect, resulting in a more uniform distribution of constituent elements, which improves the homogeneity of the passive oxide layer and mitigates the risk of galvanic coupling corrosion [25].

In conventional alloys, such as austenitic stainless steels, molybdenum alloying is being used in order to improve their resistance to pitting corrosion. It is widely recognized, that Mo addition enhances the stability of passive film in chloride solutions [25,27,28]. In the work [27] the effect of varying the molybdenum content on the corrosion resistance of $CoCrFeNiMo_x$ (where x = 0.1, 0.3 and 0.6) in NaCl solutions of 0.25 M and 1.0 M was studied. It was found that minor Mo addition (2.44 at.%) positively affects the corrosion resistance of the CoCrFeNi alloy – however, in larger quantities, when the second phase is formed, has a detrimental effect. Molybdenum influences the composition of the passivation layer, promoting the formation of $Cr_2O_3$ instead of $Cr(OH)_3$ and introducing the molybdenum oxide. An increase in $Cr_2O_3$ content of a $Cr_2O_3$ leads to the formation of a more stable and denser oxide film, while molybdenum oxides reduce the susceptibility of the alloy to pitting corrosion [25,28]. Furthermore, it was proven that Mo also contributes to a slight increase in the thickness of the passivation layer. In turn, elemental segregation caused by the formation of intermetallic phases, in the case of higher molybdenum concentrations, results in the occurrence of galvanic coupling and preferential dissolution of Cr and Mo-depleted regions [27].

Similarly, niobium can improve the corrosion resistance of the CoCrFeNi alloy by improving the protective abilities of the passivation layer [25]. In the work [29] the corrosion behaviour of the $CoCrFeNiNb_x$ alloy (x = 0.15, 0.33 and 0.5) in a 3.5% NaCl solution was studied. It was shown, that the $CoCrFeNiNb_{0.15}$ alloy characterize with the lower corrosion current density and wider passivation region – as compared to the base alloy. In turn, for alloys with the highest niobium concentration, inferior results were obtained, which was associated with elemental segregation, caused by the increase in the content of intermetallic Laves phase.

The addition of Si can improve the corrosion resistance of high-entropy alloys [25]. Yang et al. [30] investigated the influence of silicon addition on the corrosion behaviour of the $Al_{0.3}CoCrFe_{1.5}Ni$ alloy with the single-phase FCC structure. It was shown that the minor addition of silicon improves the corrosion resistance in a 3.5% NaCl solution, which was indicated by the decrease in the corrosion current density from 582 $nA/cm^2$, recorded for the base alloy, to 214 $nA/cm^2$ for $Al_{0.3}CoCrFe_{1.5}NiSi_{0.1}$. Furthermore, it was found that silicon inhibits active dissolution, which results in mitigation of the alloy susceptibility to pitting corrosion. However, a dual-phase microstructure formed in case of higher silicon content, causes galvanic coupling corrosion, resulting in deterioration of the corrosion resistance.

While, the microalloying of the conventional alloys by Mo, Nb, Si and B is well described in the literature, the effect of the higher concentration of these elements on the corrosion behaviour of HEAs is still not well described. Accordingly, the aim of this work is to compare the effect of different alloying elements, including metalloids (Si,B) and refractory elements (Mo,Nb) on the corrosion behaviour of the CoCrFeNi alloy in a chloride ion environment and their nanomechanical properties.

## 2. Materials and methods

The investigated HEAs were prepared in the form of ingots, using high purity elements (99,99 wt.%). Ingots were obtained by the induction melting method, using an NG-40 induction generator, under an argon atmosphere. The phase composition of the alloys was analysed using an X-ray diffractometer, Rigaku MiniFlex 600, equipped with a copper tube (Cu Kα, λ = 0.15406 nm) and a D/TEX strip detector. Phases were identified using dedicated PDXL2 software and the PDF-2 database. The microstructure of ingots was analysed by using scanning electron microscopy (SEM), Phenom ProX. Chemical element maps were performed using EDX spectrometer. Transmission electron microscopy (TEM) images and selected area diffraction patterns (SAED) of ingots were collected using S/TEM TITAN 80–300 FEI microscope at the operating voltage of 300 kV. The $^{57}$Fe Mössbauer transmission spectra for ingots were recorded at room temperature using an MS96 spectrometer and a linear arrangement of a $^{57}$Co:Rh source, a multichannel analyser, an absorber, and a detector. The spectrometer was calibrated at room temperature with a 30 μm thick α-Fe foil. Numerical analysis of Mössbauer spectra was performed using the MossWinn4.0i programme.

To evaluate the corrosion resistance of the investigated HEAs, electrochemical measurements were performed in the 3.5 and 5% sodium chloride solutions at a temperature of 25 °C. An Autolab 302N potentiostat, equipped with a three-electrode measuring system, was used for the study. The instrument was controlled with NOVA 1.11 software. A saturated calomel electrode (SCE) was used as a reference electrode, platinum wire was used as the counter electrode, and the material was tested as the working electrode. During electrochemical tests, changes in open circuit potential ($E_{OCP}$) were recorded, following polarization measurements in the range of –400 mV to 400 mV, with a scan rate of 1 mV/s. The corrosion potential ($E_{corr}$) and corrosion current density ($j_{corr}$) were also determined using the Tafel extrapolation method. To complete the polarization tests, electrochemical impedance spectroscopy (EIS) was performed at open-circuit potentials with an amplitude adjusted at 5 mV over a frequency range of $10^5$ - $10^{-2}$ Hz. Moreover, scanning Kelvin probe force microscopy (SKPFM) was used to determine the interaction between the phase structure and potential differences. Measurements were made on the AFM AIST-NT SmartSPM 1000 device using the ElectriTap190-G s probe with Cr/Pt coating. For the measurements, a force of 48 Nm$^{-1}$ and a resonance frequency of 190 kHz were used.

Nanomechanical tests (hardness and Young's modulus) were conducted on a NanoTest Vantage apparatus equipped with a diamond Berkovitch indenter (Synton-MDP). The apparatus was calibrated before measurements, and the diamond area function (DAF) of the indenter tip was determined by performing a series of indentations with different loads on fused silica. For the samples tested, 25 indentations were made a spacing of a load of 50 mN and with a spacing of 30 μm spacing between the indentations on a surface of 150×150 μm. The maximum load was held for 2 s, and the dwell period for drift correction was 60 s to eliminate creep of the specimen. The hardness was estimated from the indentation load-displacement curves during loading and unloading based on the Oliver-Pharr method [31]. The hardness values were calculated on equation (1):

$$H = P_{max}/A \tag{1}$$

where: $P_{max}$ is maximum load, and $A$ is the projected contact area at a specific peak load.
Furthermore, Young's modulus was calculated on the equation (2):

$$\frac{1}{E_r} = \frac{1-v^2}{E} + \frac{1-v_i^2}{E_i} \tag{2}$$

where: $E$ is Young's modulus of the sample, $\upsilon$ is Poisson's ratio of the sample, $E_i$ is Young's modulus of the indenter, $\upsilon_i$ is Poisson's ratio of the indenter. The spherical indenter with a radius of 25 μm was used to determine the stress-strain curves from nanomechanical tests [32]. A series of loads was applied between 1 and 500 mN during 30 cycles.

## 3. Results and discussion

According to the complex chemical composition, the prepared ingots were described based on the analysis of XRD patterns, EDX 2D maps, TEM images, and SAED analysis. Analysis of the $^{57}$Fe Mössbauer transmission spectra supported the obtained results. First, the chemical elements distribution was analysed on the EDX spectra (**Fig. 1**). As can be seen, the segregation of the chemical elements appears for all samples, suggesting the coexistence of a few phases. In the case of CoCrFeNiMo alloy (**Fig. 1a**), only the chromium is nearly uniformly distributed, while the distribution of the Co and Fe is similar to the distribution of Ni. However, three different regions can be observed. The first relates to areas rich in Ni, Fe, Co, Cr, and poor in Mo. The second region is rich in Mo, where Co, Cr, and Fe can also be observed. Interestingly, the presence of the regions in which both Mo and Ni are presented can also be seen in **Fig. 1a**. High segregation of chemical elements was also observed for the CoCrFeNiNb (**Fig. 1b**) and CoCrFeNiB (**Fig. 1c**) alloys. In the CoCrFeNiNb alloy, only Co was nearly uniformly distributed, while segregation appears between Nb and Cr, Fe, and Ni, i.e., the presence of two areas rich and poor in Nb can be observed. The more complicated structure was observed for alloy with boron. The 2D EDX maps present regions rich in Cr, poor in Ni and Co, and rich in Cr and Fe with the presence of Co. Nickel is mainly present in regions rich in Fe and Co. The nearly uniform distribution of all chemical elements was observed for the CoCrFeNiSi alloy (**Fig. 1d**). Segregation was observed only locally, resulting in areas rich in Ni and Si elements.

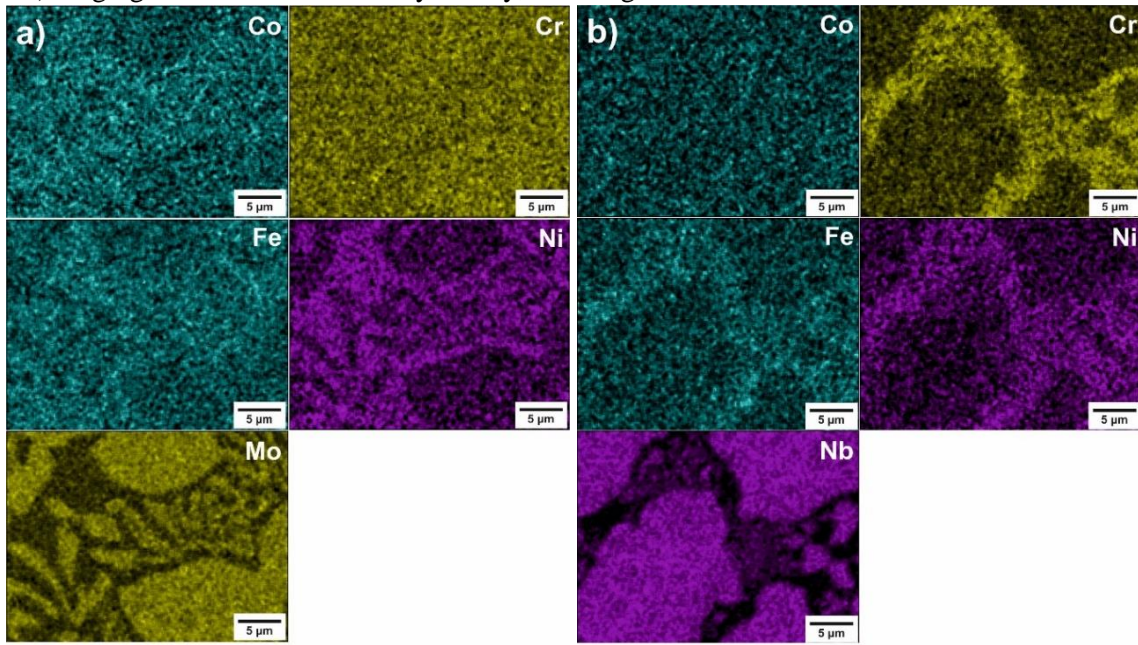

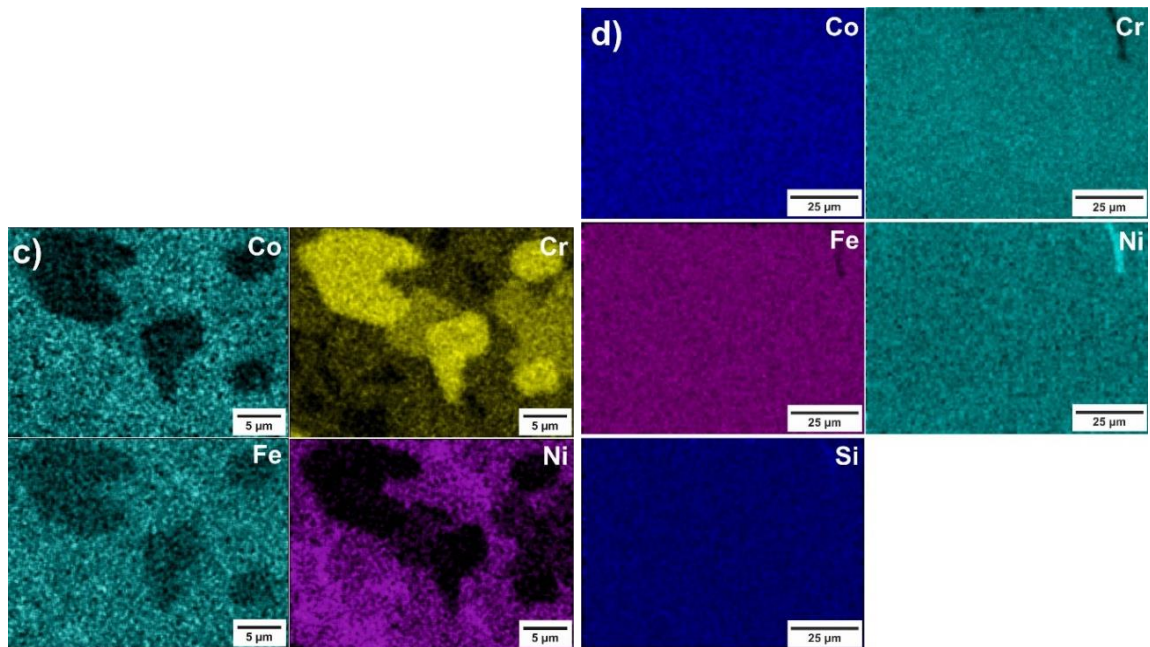

**Fig.1.** 2D EDX maps of (a) CoCrFeNiMo, (b) CoCrFeNiNb, (c) CoCrFeNiB and (d) CoCrFeNiSi HEAs in form of ingots

The analysis of the XRD patterns (**Fig. 2**) confirms the results of the 2D EDX maps. Two phases were observed for the CoCeFeNiNb alloy. The first of them is related to the FCC phase rich in Fe, Ni, Cr, and Co (space group: $Fm\bar{3}m$) and the second one with intermetallic phase marked as $(TM)_2Nb$ (hexagonal phase; space group: $P6_3/mmc$), where TM can be related to the presence of Co, Cr, Fe and Ni. In the case of the CoCrFeNiMo alloy, three phases were identified. One of them, marked as FCC (cubic phase; space group: $Fm\bar{3}m$) is related to the crystalizing of a solid solution rich in Fe, Ni, Cr and Co. The two other phases (marked together in the XRD pattern as Cr-Mo-TM) are related to intermetallic phases rich in Mo – $Cr_9Mo_{21}Ni_{20}$ (orthorhombic phase; space group: Pbnm) and tetragonal phase (space group: $P4_2/mnm$) $Cr_{0.8}Mo_{0.4}Ni_{0.8}$, in which Fe and Co substitute Ni atoms. Moreover, as was presented above, one of these phases should be rich in Ni, while the second in Fe and Co. Four different phases were identified for alloys containing boron. The presence of this complex phase structure remains in line with the findings from the analysis of 2D EDX maps. The observed regions, rich in Cr and poor in Ni and Co, can be attributed to the orthorhombic $Cr_2B$ phase (space group: Fddd). The areas rich in Cr, Fe, and Co can be related to the crystallization of $(Co,Cr)B$ phase (tetragonal phase, space group: I4/mcm), in which Fe and Ni can substitute Co. The regions in which the presence of Fe, Co, Ni, and Cr was observed can be related to the FCC phase (cubic phase; space group: $Fm\bar{3}m$) and regions poor in Cr to the intermetallic phase $Fe(Co,Ni)_2B$ (orthorhombic phase; space group: Pnma). In the case of the CoCrFeNiSi alloy, two intermetallic phases were observed. The first, marked as $TM_{3.921}Si_{1.079}$, is a cubic phase (space group: P213), in which TM can be Fe, Ni, Cr, and Co, according to the EDX analysis. The second, also observed under the analysis of 2D EDX maps, is related to the formation of the $Ni_2Si$ orthorhombic phase (space group: Pbnm).

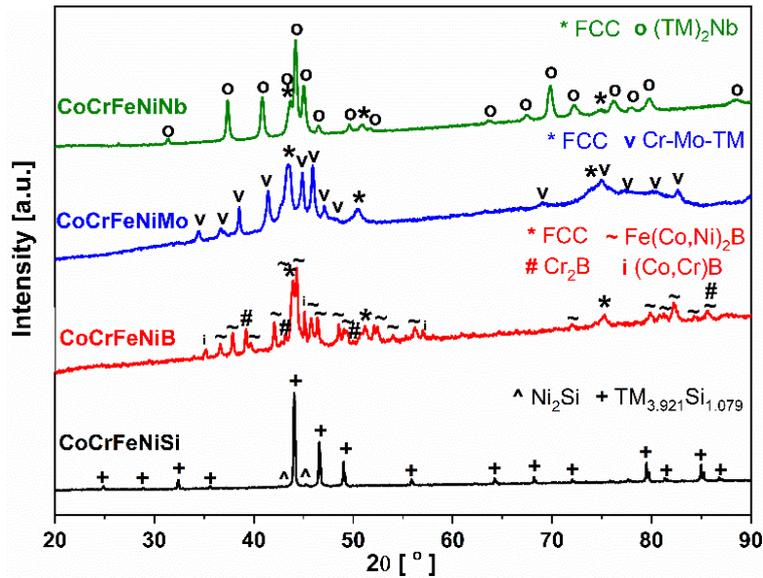

**Fig.2.** XRD patterns of CoCrFeNiX (X=Mo,Nb,Si,B) HEAs in a form of ingots

The $^{57}$Fe Mössbauer transmission spectra were recorded and analyzed to confirm the XRD analysis. The room-temperature Mössbauer spectra of the CoCrFeNiX (X=Mo,Nb,Si,B) alloys in the form of an ingot, together with the fitted components, are presented in **Fig. 3**. The hyperfine parameters for all fitted components are listed in **Table 1**. The Mössbauer spectrum for the CoCrFeNiNb alloy (**Fig. 3a**) was fitted with two quadrupole doublets. The hyperfine parameters of the main doublet indicate the presence of Fe in the TM$_2$Nb, the hexagonal structure [33]. Slight differences in the hyperfine parameters may indicate the substitution of Fe atoms for other alloy elements. The second doublet visible in the Mössbauer spectrum of this alloy (**Fig. 3a**) is associated, as indicated by the XRD results, with Fe atoms in the FCC structure. A small negative value of the isomer shift (**Table 1**) for this doublet is typical for this structure that contains Fe, Ni, and Cr [34]. The quadrupole splitting indicates a high disorder of this phase, probably resulting from the presence of Co and Nb atoms.

The Mössbauer spectrum of the CoCrFeNiMo alloy (**Fig. 3b**) was fitted with a sextet, a quadrupole doublet, and a single line. This sextet should be assigned to Fe atoms in the FCC structure, and the value of the hyperfine magnetic field (**Table 1**) indicates that it is a phase rich in Fe that also contains Ni or Co atoms [35]. The single line visible in the Mössbauer spectrum of this alloy will also be related to Fe in the FCC structure. The negative value of the isomer shift and the absence of the hyperfine magnetic field indicate that Fe in its surroundings, in addition to Ni atoms, also contains Cr atoms. As suggested by the XRD results (**Fig. 2**), the quadrupole doublet is associated with the some Cr-Mo-TM (TM=Co,Fe,Ni) structure [36].

Two sextets visible in the Mössbauer spectrum of the CoCrFeNiB alloy (**Fig. 3c**) indicate the presence of the Fe$_2$B phase [37,38]. The hyperfine magnetic field values and the line widths of these sextets (**Table 1**) indicate a high disorder of this structure and the substitution of Fe atoms by other atoms, such as Ni, Co, or Cr. However, the main component on this spectrum is a quadrupole doublet. The hyperfine parameters of this doublet differ from those calculated for the paramagnetic components observed for the CoCrFeNiNb or CoCrFeNiMo alloys. In our opinion, this doublet might be referred to Fe atoms in the FCC-(Fe,Ni) structure, with more than ten Fe atoms as closest neighbours [38]. The high value of quadrupole splitting for this doublet is probably due to the substitution of Fe or Ni atoms by, e.g. Co, which causes distortion of the crystal lattice.

The main doublet visible in the Mössbauer spectrum of the CoCrFeNiSi alloy (**Fig. 3d**) is associated with Fe atoms in the Fe-Si system [39]. The XRD results show (**Fig. 2**) that this phase will be TM$_{.921}$Si$_{1.079}$, in which TM can be Fe, Ni, Cr and Co. The hyperfine parameters of the second doublet are very similar to those obtained for the doublet in CoCrFeNiB alloy. However, these parameters also agree with the values stated for amorphous FeSi$_2$ [39] or FeSi, where Fe is substituted by Ni [40]. Considering also the XRD results, we can assume that this component is related to Ni$_2$Si, where Fe replaces Ni.

**Table 1.** Mössbauer hyperfine parameters of the investigated the CoCrFeNiX (X=Mo,Nb,Si,B) high entropy alloys in a form of ingot; *IS* – isomer shift, *QS* – quadrupole splitting, $B_{hf}$ – hyperfine magnetic field, *G* – full line width at half maximum, *A* – area fraction of subspectra; TM = Fe, Co or Ni. Estimated errors are ±0.01 mm/s for *IS* and *G*, for *QS* is ±0.02 mm/s and ±0.1 for $B_{hf}$

| Sample | IS [mm/s] | QS [mm/s] | $B_{hf}$ [T] | G [mm/s] | A [%] | Interpretation |
|---|---|---|---|---|---|---|
| CoCrFeNiNb | -0.25 | 0.33 | - | 0.29 | 73 | (TM)$_2$Nb |
|  | -0.06 | 0.22 | - | 0.29 | 27 | FCC-Fe(Cr,Ni) |
| CoCrFeNiMo | -0.07 | 0.37 | - | 0.34 | 59 | Cr-Mo-TM |
|  | -0.09 | - | - | 0.34 | 30 | FCC-Fe(Cr,Ni) |
|  | 0.01 | 0.00 | 33.3 | 0.40 | 11 | FCC-Fe(Ni,Co) |
| CoCrFeNiB | 0.10 | 0.50 | - | 0.35 | 80 | FCC-Fe(Cr,Ni) |
|  | 0.03 | 0.00 | 27.1 | 0.58 | 12 | Fe(Co,Ni)$_2$B |
|  | 0.02 | 0.00 | 22.3 | 0.58 | 8 |  |
| CoCrFeNiSi | 0.20 | 0.50 | - | 0.28 | 63 | TM$_{3.921}$Si$_{1.079}$ |
|  | 0.08 | 0.46 | - | 0.28 | 27 | Ni(Fe)$_2$Si |

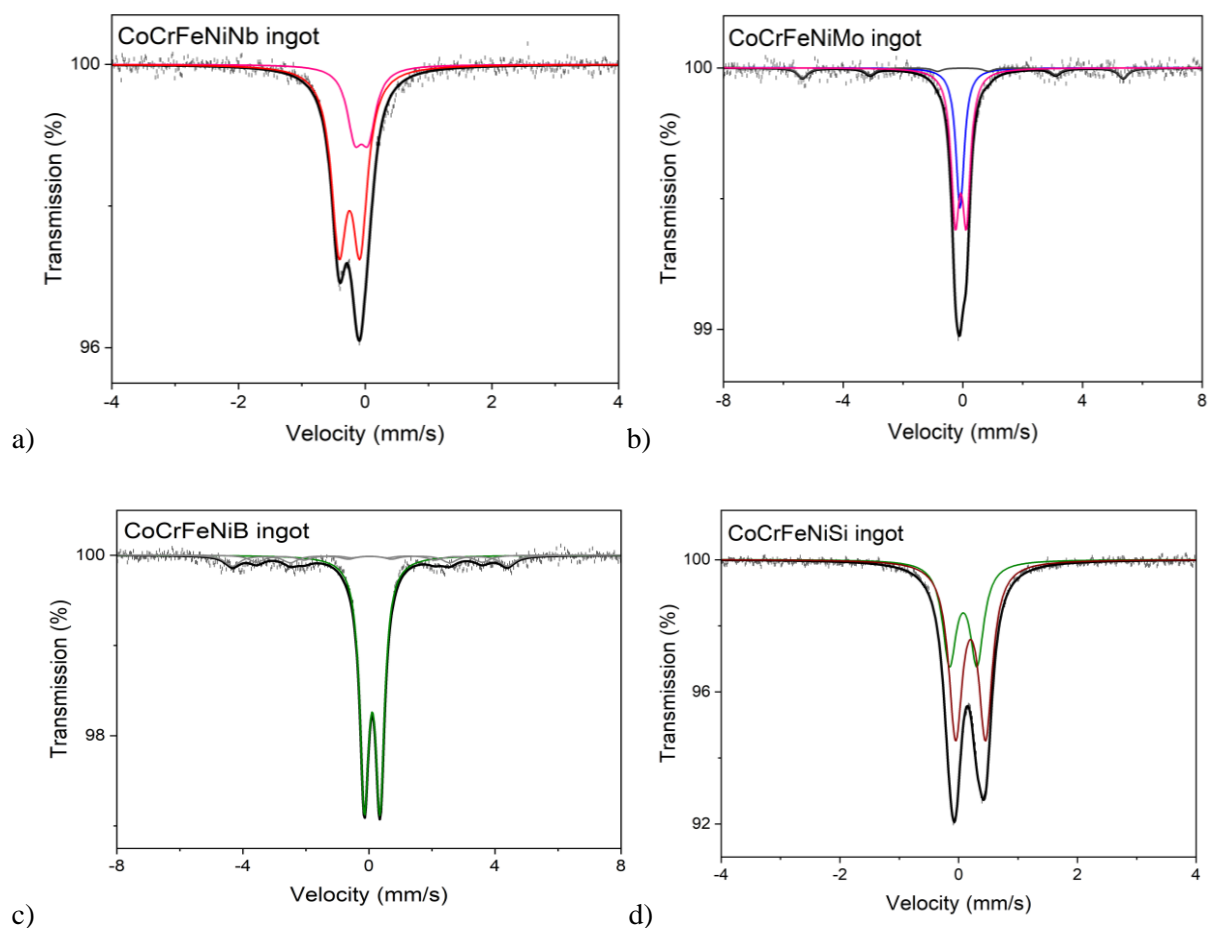

**Fig.3.** Mössbauer spectra of CoCrFeNiX (X=Mo,Nb,Si,B) HEAs in a form of ingots

In addition, to confirm the presence of the borides in the CoCrFeNiB alloy, analysis of the TEM images and SAED patterns was performed. The SAED pattern is presented in **Fig. 4a**. The expected coexistence of different phases was confirmed, and the results of the analysis are listed in **Table 2**. According to multiphase structure of the alloys, most of the measured d-spacing can be related to more than one phase. Analysis confirms the phases identified in the XRD pattern, that is,

Fe(TM)$_2$B, FCC, Cr$_2$B, and (Co,Cr)B. Moreover, the FCC phase and Fe(Co,Ni)$_2$B phase were also identified in the TEM image (**Fig. 4b**). The presence of the FCC phase confirms the analysis presented in **Fig. 4c**, while the presence of the Fe(Co,Ni)$_2$B phase was confirmed by analysis of the Fast Fourier Transform (FFT) area (**Fig. 4d**) and revealed in **Fig. 4e**. The high-resolution TEM images and corresponding SAED patterns for the remaining alloys were also recorded and presented in the **Fig 5**. In the case of the CoCrFeNiSi alloy, analysis of the obtained SAED pattern (**Fig. 5b**) indicated the presence of the TM$_{3.921}$Si$_{1.079}$ phase (TM=Fe). The lattice images of the interdendrite matrix of the CoCrFeNiMo and CoCrFeNiNb alloys were shown in **Figs. 5c** and **5e**, while corresponding SAED patterns are presented in **Figs. 5d** and **5f**. In both cases, analysis of the diffraction patterns further confirmed the occurrence of the FCC solid solution phase.

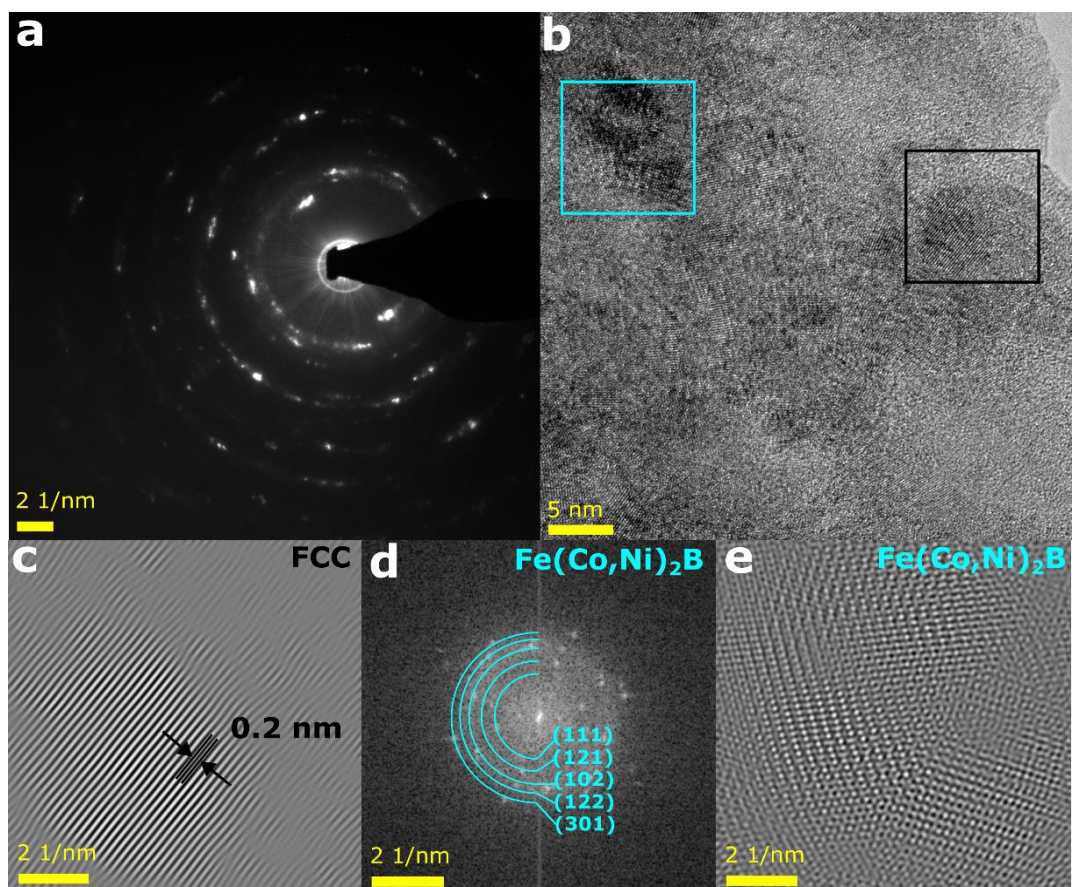

**Fig.4.** SAED pattern (a) and TEM image (b) of CoCrFeNiB HEAs in the form of an ingot with IFFT (c) from the black-marked area on (b) and FFT (d) from the blue-marked area on (b) and corresponding IFFT (e)

**Table 2.** Analysis results of the SAED patterns together with the corresponding possible phases and related Miller indices.

| Measured d-spacing [nm] | Possible phases |
|---|---|
| 0.260 | (Co,Cr)B (200); Fe(Co,Ni)$_2$B (200) |
| 0.205 | FCC (111); Fe(Co,Ni)$_2$B (102); (Co,Cr)B (211) |
| 0.190 | Fe(Co,Ni)$_2$B (221) |
| 0.165 | Fe(Co,Ni)$_2$B (212); (Co,Cr)B (202); Cr$_2$B (440) |
| 0.131 | Fe(Co,Ni)$_2$B (123); Cr$_2$B (480) |
| 0.122 | Fe(Co,Ni)$_2$B (420); (Co,Cr)B (330); Cr$_2$B (313) |
| 0.110 | Fe(Co,Ni)$_2$B (431); (Co,Cr)B (402) |

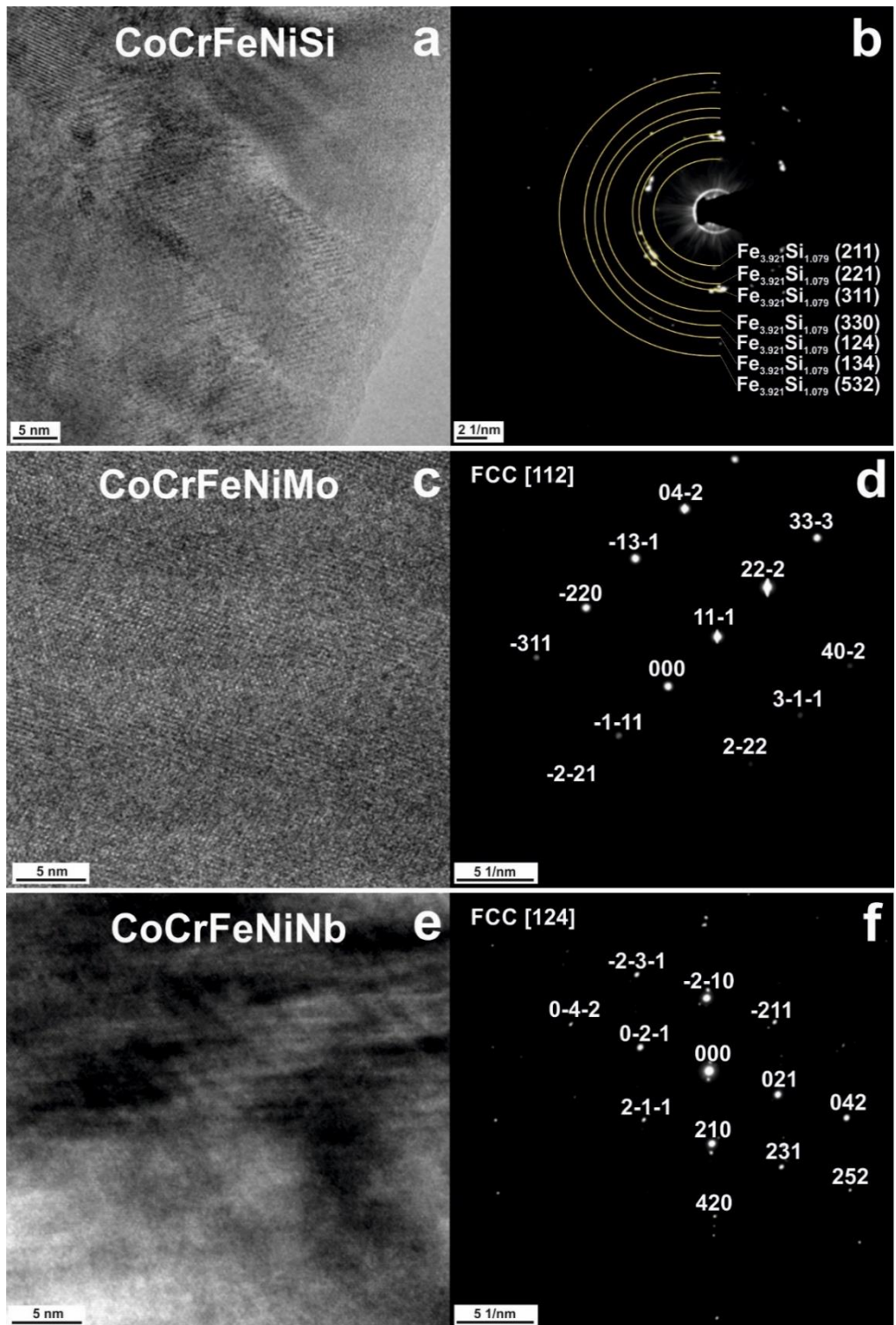

**Fig.5.** TEM image (a,c,e) and SAED pattern (b,d,f) of CoCrFeNiMo, CoCrFeNiSi and CoCrFeNiNb HEAs in a form of ingots

Finally, the microstructure of studied alloys was revealed based on the previous analysis and is presented in **Fig. 6**. The SEM image confirms the nearly one-phase structure of the CoCrFeNiSi alloys related to the crystallization of the $TM_{3.921}Si_{1.079}$ phase (**Fig. 6d**).

SEM images of CoCrFeNiMo (**Fig. 6a**) and CoCrFeNiNb (**Fig. 6b**) alloys suggest their nearly dual-phase microstructure. Both alloys are characterized by dendritic morphology, with different types of structures present in the interdendritic regions. Significantly higher volume content of dendrites can be observed for the latter alloy. In the case of the CoCrFeNiNb alloy, the light-grey dendrites are enriched with niobium; therefore, they can be related to the (TM)$_2$Nb phase. As niobium has the highest melting point than the rest of constituent elements, the (TM)$_2$Nb phase crystallizes first, forming the dendrites during solidification [42,43]. The darker matrix is enriched with chromium, iron, and nickel, whereas depleted of niobium consequently can be identified as the FCC solid solution. There are different types of structures in the interdendritic regions: larger, bright precipitates constituted of the (TM)$_2$Nb phase followed by eutectic colonies (**Fig. 6b**). Similarly, Cr-Mo-TM phases forms the dendrites in the CoCrFeNiMo alloy (**Fig. 6a**). Furthermore, according to the analysis of the XRD patterns, the crystallization of two different phases with Mo (Cr$_9$Mo$_{21}$Ni$_{20}$ and Cr$_{0.8}$Mo$_{0.4}$Ni$_{0.8}$) appears, while the Cr$_{0.8}$Mo$_{0.4}$Ni$_{0.8}$ phase probably crystallizes at Cr$_9$Mo$_{21}$Ni$_{20}$ grains boundaries. The darker matrix corresponds to the solid FCC solution, being depleted of molybdenum, as a result of its limited solid solubility [28]. In the interdendritic regions, eutectic structures consisting of alternating lamellae of Cr-Mo-TM and FCC solid solution can be observed.

The complex multiphase structure previously discussed of CoCrFeNiB alloy can also be visible in the SEM image (**Fig. 6c**). Scattered light-grey precipitates of elongated shape can be related to the crystallization of Fe(Co,Ni)$_2$B phase. In turn, the grey areas are related to the formation of the FCC solid solution, which, according to the Mössbauer spectra analysis, is the main phase in this alloy. Interestingly, dark grey precipitates of similar morphology are related to the crystallization of Cr-rich Cr$_2$B and Co and Cr-rich (Co,Cr)B borides, and these phases can only be distinguished based on the EDX analysis.

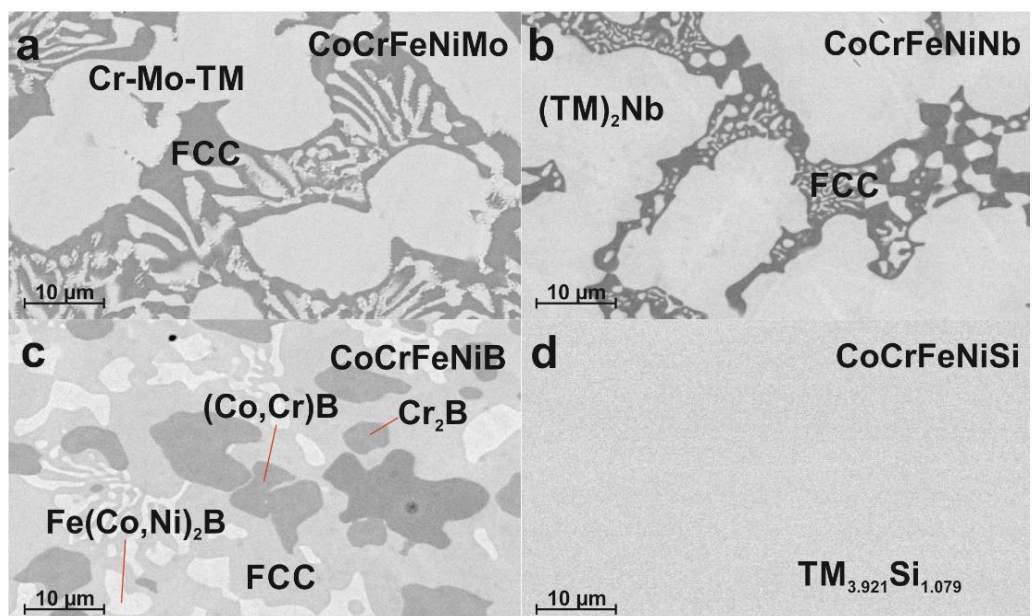

**Fig.6.** SEM microstructures of CoCrFeNiX (X=Mo,Nb,Si,B) HEAs

To compare the effect of different alloying elements on the corrosion behaviour of the CoCrFeNi alloy, electrochemical tests were carried out, in sodium chloride solutions of two different concentrations, at 25°C. Changes as open-circuit potential in a function of time, for measurements in milder 3.5% NaCl solution, were presented in **Fig. 7a**. The most positive open-circuit potential values were obtained for the CoCrFeNiNb alloys (0.026 V) and CoCrFeNiSi (0.005 V), while the CoCrFeNiB alloy exhibit the lowest $E_{OCP}$ value (-0.150 V).

**Fig. 7b** shows the recorded polarization curves, in 3.5% NaCl solution, at 25°C. The parameters obtained as a result of the measurements performed are summarized in **Table 3**. Based on the tests conducted, it can be stated that the best corrosion resistance in this environment characterize the CoCrFeNiSi alloy, as indicated by the lowest value of the corrosion current density (1.77 µA/cm$^2$) and the highest value of the polarization resistance (75.05 kΩcm$^2$). Noticeably, the polarization curve obtained for this alloy displays the anodic peak, contrastingly to the rest alloys measured in that environment. The alloy with Si addition also shows the most positive corrosion potential (-

0.068 V), whereas the least favourable corrosion potential value characterises the CoCrFeNiMo alloy (-0.222 V).

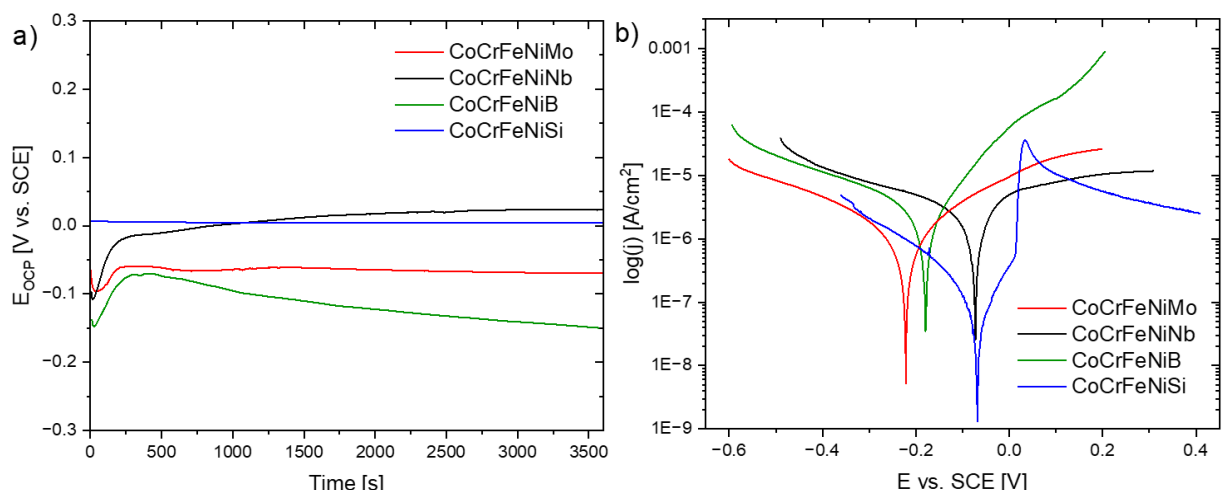

**Fig.7.** Changes of open-circuit potential (a) and polarization curves (b) of CoCrFeNiX (X=Mo,Nb,Si,B) HEAs in 3.5% NaCl solution at 25 °C

**Table 3.** Results of the electrochemical tests conducted in the 3.5% NaCl solution, at 25°C

| Sample | $E_{corr}$ [V] (±0.01) | $j_{corr}$ [µA/cm$^2$] (±0.1) | $R_p$ [kΩcm$^2$] (±0.1) |
|---|---|---|---|
| CoCrFeNiMo | -0.222 | 50.97 | 7.71 |
| CoCrFeNiNb | -0.073 | 62.43 | 5.56 |
| CoCrFeNiB | -0.179 | 31.97 | 4.92 |
| CoCrFeNiSi | -0.068 | 1.77 | 75.05 |

To determine the corrosion behaviour of the alloys in more demanding corrosion environment, additional tests were performed in 5% NaCl solution, at 25°C. **Fig. 8a** presents the changes as the open-circuit potential in a function of time, whereas the recorded polarisation curves were shown on **Fig. 8b**. The quantitative parameters of the corrosion potential, the corrosion current density and the polarization resistance were summarized in **Table 4**.

Compared to the results obtained for measurements in a 3.5% NaCl solution, significant differences can be observed. The open-circuit potentials shifted towards more negative values for all compositions. In this case, the best potential characterize the CoCrFeNiSi alloy (-0.112 V), while the least favourable was obtained for CoCrFeNiB (-0.290 V). A similar tendency can be observed in terms of the corrosion potentials, particularly pronounced for the CoCrFeNiB alloy. Interestingly, there is a decrease in corrosion current density values obtained for CoCrFeNiNb, CoCrFeNiSi and CoCrFeNiB alloys. Significant enhancement can be noticed especially for the CoCrFeNiNb alloy, which exhibit two orders of magnitude lower corrosion current density and correspondingly higher polarization resistance. However, similarly to the previous environment, the highest corrosion resistance in the 5% NaCl solution, which exhibits the lowest corrosion current density (0.24 µA/cm$^2$) and the highest polarization resistance (331.99 kΩ). On the contrary, the worst corrosion behaviour characterises the CoCrFeNiMo alloy, which was indicated by the highest corrosion current density (59.2 µA/cm$^2$) and the lowest resistance to polarization (6.17 kΩ). Similarly, the most positive value can be observed for the CoCrFeNiSi alloy (-0.136 V), whereas the least favourable result was noticed for the CoCrFeNiB (-0.339 V).

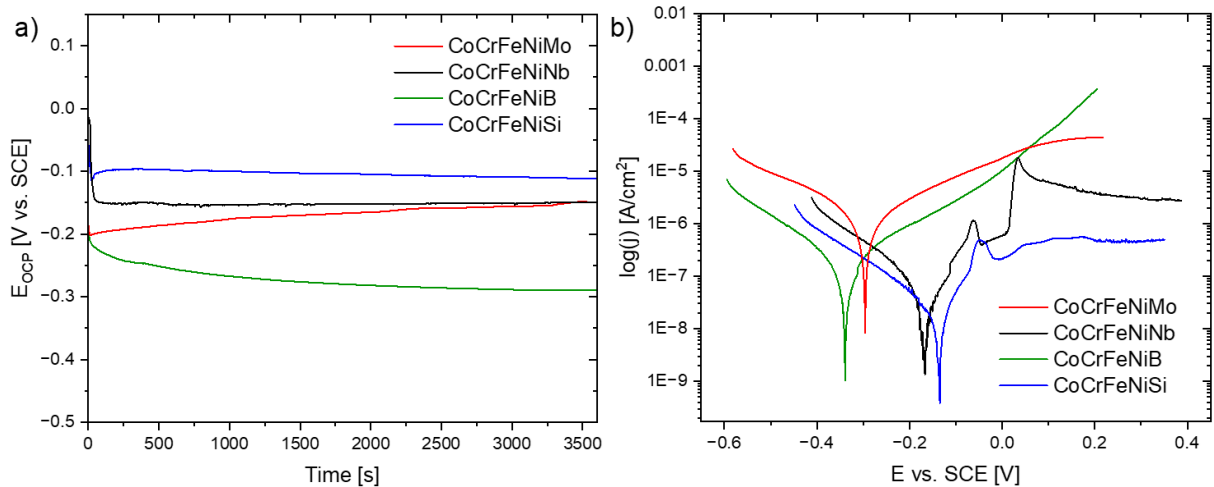

**Fig.8.** Changes of open-circuit potential (a) and polarization curves (b) of CoCrFeNiX (X=Mo,Nb,Si,B) HEAs in 5% NaCl solution at 25 °C

**Table 4.** Results of the electrochemical tests conducted in the 5% NaCl solution, at 25°C

| Sample | $E_{corr}$ [mV] (±0.01) | $j_{corr}$ [µA/cm$^2$] (±0.01) | $R_p$ [kΩcm$^2$] (±0.01) |
|---|---|---|---|
| CoCrFeNiMo | -0.296 | 59.20 | 6.17 |
| CoCrFeNiNb | -0.168 | 0.63 | 135.11 |
| CoCrFeNiB | -0.339 | 2.07 | 66.21 |
| CoCrFeNiSi | -0.136 | 0.24 | 331.99 |

The microstructure is one of the main factors that affect the corrosion resistance of the tested alloys. The presence of intermetallic phases can result in elemental segregation, causing the formation of distinct zones with varying potentials, leading to the appearance of galvanic coupling corrosion [25,44]. In case of CoCrFeNiMo, the intense segregation of the elements, clearly visible on the 2D EDX maps (**Fig. 1a**), related the presence of (Mo,Ni)-rich $Cr_{0.8}Mo_{0.4}Ni_{0.8}$ phase, and especially, large volume content of Mo-rich $Cr_9Mo_{21}Ni_{20}$ phase, contributed to enhanced corrosion susceptibility. It was previously reported that alloying with Mo of higher concentrations can deteriorate the corrosion resistance of the CoCrFeNi alloy [28,45]. Similarly, in the case of the CoCrFeNiNb alloy, pronounced segregation of Nb and Cr can be observed, although in this case its negative effect on the corrosion resistance is less unambiguous. Previous work confirms the beneficial effect of adding niobium on the passivation ability of CoCrFeNi alloy, however the adverse effect of intermetallic phase formation was also indicated [25,29]. The inverse segregation of chromium and niobium, which can be observed in the 2D EDX maps (**Fig. 1b**) lead to formation of Cr-enriched regions and Nb-enriched regions – simultaneously, both niobium and chromium can be conducive to high stability and protective abilities of passivation layer. Moreover, the complex phase structure and resulting heterogeneous distribution of elements, especially pronounced for chromium, resulted in the diminished passivation ability of the CoCrFeNiB alloy. In contrast, the superior corrosion resistance of the CoCrFeNiSi alloy (**Fig. 1d**) may be related to its relatively homogeneous elemental distribution.

The corrosion behaviour of the CoCrFeNiNb$_x$ (x = 0, 0.2, 0.4, 0.6 and 1.0), in 1M NaCl solution at 30°C, was analysed in publication [46]. In that study, the CoCrFeNiNb$_{0.2}$ and CoCrFeNiNb alloys exhibited similar values of the corrosion current density of approximately 12 µA/cm$^2$, noticeably more favourable than the values obtained in this work for equimolar CoCrFeNiNb during measurements in 3.5% NaCl solution. Simultaneously, much lower $j_{corr}$ values were obtained for tests conducted in more concentrated saline solution.

Other works also describe the effect of Si on the corrosion behaviour of HEAs, however, with different chemical compositions and Si content. Yang et al. [30] studied the dependence of corrosion resistance on the silicon concentration in $Al_{0.2}CoCrFe_{1.5}NiSi_x$ alloy (where x = 0, 0.1, 0.2, 0.3) in the 3.5% NaCl solution, at 25°C. In that paper, the best corrosion resistance was demonstrated by $Al_{0.2}CoCrFeNiSi_{0.1}$, exhibiting the corrosion current density of approximately 0.21 µA/cm$^2$. Concurrently, increased Si content caused the precipitation of secondary phase, contributing to more heterogenous elemental distribution and deterioration of the corrosion resistance. However, in the case of the CoCrFeNiSi described in this work, the corrosion behaviour was different because of the more homogeneous elemental distribution. In article [47], corrosion behaviour of $CoCrFeNiB_x$ (where x = 0, 0.01, 0.05, 0.1 and 0.5) prepared using high-energy ball milling was investigated. Similar values of corrosion current density were obtained for most alloys, ranging around 2.3-2.9 µA/cm$^2$. The less advantageous corrosion behaviour of the CoCrFeNiB alloy can be assigned to a more heterogeneous distribution of chromium, related to a larger volume content of borides.

In order to further investigate the passivation of the CoCrFeNi-(Mo,Nb,Si,B) alloys, electrochemical impedance spectroscopy measurements were performed. Nyquist plots (**Figs. 9a** and **9b**) indicated substantial differences in the corrosion behaviour of the studied alloys. The CoCrFeNiSi and CoCrFeNiNb characterize with the presence of one capacitive semicircle; however, in case of the latter alloy, it merges with straight line in low frequencies. This characteristic indicates that the kinetics of electrode reactions was controlled by diffusion [48]. In turn, the spectra recorded for the CoCrFeNiMo and CoCrFeNiB alloys show two poorly separated capacitive semicircles, suggesting the porosity of the passive film formed on the alloy surface [49].

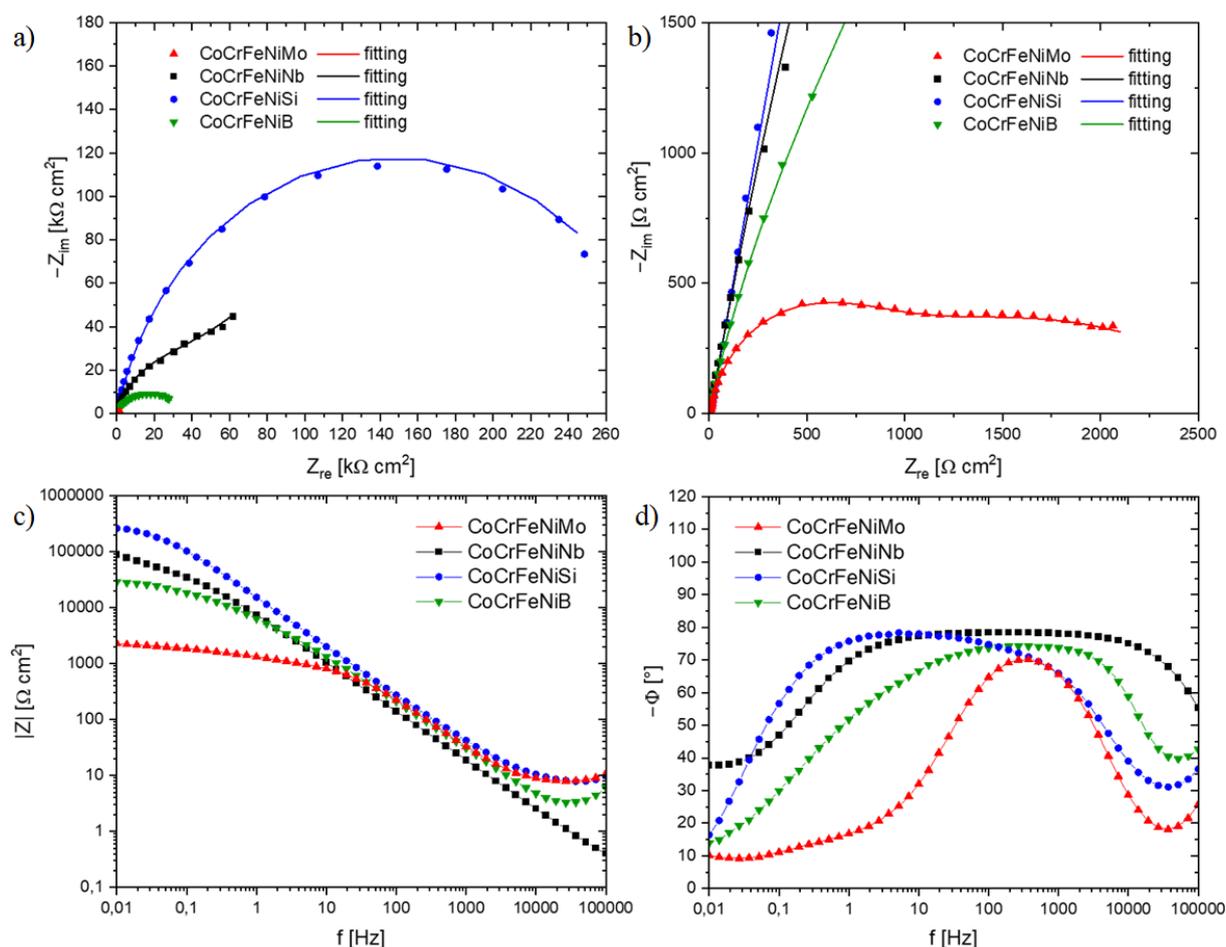

**Fig.9.** Experimental EIS spectra: Nyquist plots (a) – enlarged initial section of diagram shown at (b), Bode modulus plots (c), Bode phase angle plots (d) for CoCrFeNiX (X=Mo,Nb,Si,B) alloys in form of ingots in 3.5% NaCl solution, at 25 °C.

Usually, the semicircles in the high frequency range are associated with the charge transfer at the electrode interface, and their diameter reflects the impedance of the passive film [48,50]. The larger diameter of the semicircle indicates higher impedance and consequently better corrosion resistance of the passive film. Therefore, the recorded Nyquist plots further confirm the highest corrosion resistance of the CoCrFeNiSi alloy. Simultaneously, the least advantageous corrosion behaviour characterize the CoCrFeNiMo alloy. The Bode plot of this alloy displays the minimum phase angle at -75°, with small broadening at the medium frequencies, which indicate that no stable passivation layer has formed on the alloy surface. In turn, in case of the CoCrFeNiNb alloy, the Bode diagram displays the phase angle close to -80° in wide range of frequencies, indicating better protective abilities of the passive film formed on its surface [51]. Also the CoCrFeNiSi alloy exhibit phase angle approaching to -80°, with substantial broadening at medium frequencies, evidencing high stability of its passivation layer.

Three suitable electric equivalent circuits were established to analyse the obtained impedance spectra (**Fig. 10**). The simplest circuit, shown in **Fig. 10b**, used for fitting the spectra obtained for CoCrFeNiSi alloy consist of solution resistance ($R_s$), passive film resistance ($R_p$) and the constant phase element (CPE), which represents the capacitance of electrical double-layer at the electrode surface. The latter element substituted the pure capacitance to account the frequency dispersion effect, which can arise from the inhomogeneity of the passive film or surface roughness of the electrode surface [29]. The impedance of the constant phase element can be determined employing the following equation [48, 52]:

$$Z(\omega) = Z_0(i\omega)^{-n} \qquad (3)$$

where: $Z_0$ is the capacitance of the parameter related to the electrode capacitance [F · cm$^{-2}$ · s$^{n-1}$]; $\omega$ represents the angular frequency (rad/s) and $n$ is the constant phase exponent. The CPE can represent the pure capacitance, when the $n = 1$, the Warburg impedance ($n = 0.5$) or can be purely resistive, when $n = 0$ [52]. In case of the CoCrFeNiNb alloy the proposed electric equivalent circuit (EEC) (**Fig. 10a**) displays also the Warburg impedance to account the diffusion-related effects ($W$) [48, 53]. In turn, for fitting the spectra of the CoCrFeNiMo and CoCrFeNiB alloys, circuit with two time constants was used (**Fig. 10c**), assuming the formation of porous passivation layer [52]. The EEC used is comprised of solution resistance ($R_s$), passive film resistance ($R_1$), charge transfer resistance ($R_2$), constant phase elements representing the passive film (CPE$_1$) and electrical double-layer capacitance (CPE$_2$) [54].

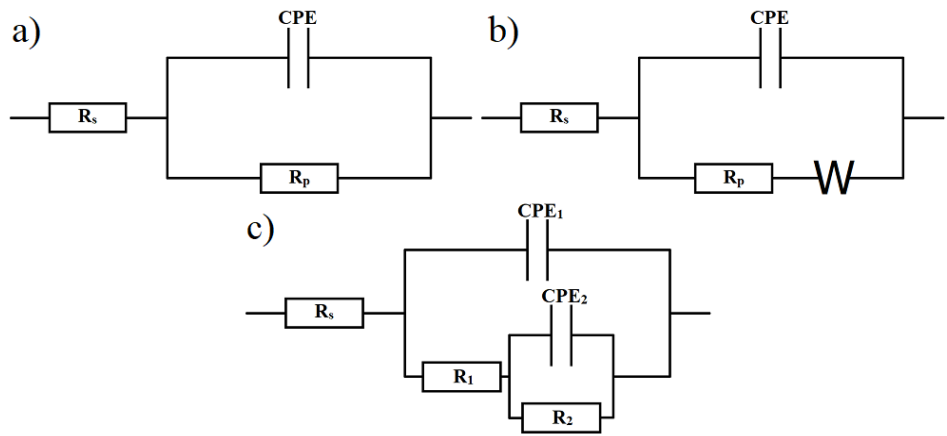

**Fig.10.** Proposed equivalent electric circuits for CoCrFeNiSi alloy (a), CoCrFeNiNb alloy (b), CoCrFeNiMo and CoCrFeNiB alloys (c) in 3.5% NaCl solution at 25 °C

The values of parameters fitted for CoCrFeNiNb and CoCrFeNiSi are shown in **Table 5**, whereas the fitting results for the alloys remaining were presented in **Table 6**. The polarization resistance of the CoCrFeNiB and CoCrFeNiMo alloys can be approximated as sum of the passive film resistance and charge transfer resistance [55]. The highest value of polarization resistance characterize the CoCrFeNiSi alloy (296.1 kΩ), which is in consistency with the results of potentiodynamic polarization measurements. Concurrently, in case of the CoCrFeNiMo, two orders of magnitude lower value of polarization resistance was noticed (5.6 kΩ). The significant value of double layer capacitance can be noticed in case of CoCrFeNiMo alloy. Resulting higher surface charge density, contributes to enhanced movement of ions across the double layer, therefore accelerating the corrosion processes [54].

**Table 5.** The fitted electrochemical parameters for impedance data of the CoCrFeNiNb and CoCrFeNiSi in the as-cast state in 3.5% NaCl solution, at 25°C.

| Sample | $R_s$ [$\Omega cm^2$] | CPE [$\mu\Omega^{-1}cm^{-2}s^n$] | $n$ | $R_p$ [$k\Omega cm^2$] | W [$\mu\Omega^{-1}cm^{-2}s^n$] |
|---|---|---|---|---|---|
| CoCrFeNiNb | 1.9 | 28.1 | 0.85 | 48.0 | 62.6 |
| CoCrFeNiSi | 6.8 | 13.7 | 0.86 | 296.1 | - |

**Table 6.** The fitted electrochemical parameters for impedance data of the CoCrFeNiMo and CoCrFeNiB alloys in the as-cast state in 3.5% NaCl solution, at 25°C.

| Sample | $R_s$ [$\Omega cm^2$] | $CPE_1$ [$\mu\Omega^{-1}cm^{-2}s^n$] | $n_1$ | $R_1$ [$k\Omega cm^2$] | $CPE_2$ [$\mu\Omega^{-1}cm^{-2}s^n$] | $n_2$ | $R_2$ [$k\Omega cm^2$] |
|---|---|---|---|---|---|---|---|
| CoCrFeNiMo | 8.9 | 11.33 | 0.91 | 0.6 | 484.7 | 0.37 | 2.21 |
| CoCrFeNiB | 2.1 | 20.0 | 0.85 | 3.5 | 37.1 | 0.46 | 34.8 |

The previous works indicate the bilayer structure of the passive film formed on the CoCrFeNiMo$_x$ alloy surface [45,56]. It was found, that the inner layer was composed primarily of oxides, with high concentration of $Cr_2O_3$, and therefore is decisive in terms of passive film protective abilities. Also the minor amount of $MoO_4$ was present. Concurrently, the outer layer exhibits higher concentration of hydroxides, mainly of chromium and iron, followingly with $MoO_3$ [45,56]. The presence of Mo oxides incorporated in the passive film can contribute to its higher stability, as the molybdenum causes deprotonation of $Cr(OH)_3$, translating into generation of more $Cr_2O_3$ [28]. Simultaneously, elemental segregation, related to the formation of Cr-Mo-TM intermetallic phases, may cause the nonuniform chromium distribution in the passivation layer, thereby negatively affecting its corrosion resistance. Moreover, in case of high molybdenum concentrations its positive effect on the passive film declines, resulting in formation of more defective layer with lower $Cr_2O_3/Cr(OH)_3$ ratio [45].

Earlier works revealed the beneficial effect of minor niobium addition on protective abilities of the CoCrFeNi alloy passive film [25,29]. Similarly as in case of molybdenum, the addition of niobium results in formation of more $Cr_2O_3$ at the expense of $Cr(OH)_3$, enhancing the passive film compactness. Concurrently, generation of $Nb_2O_5$ further enhance the stability of the passivation layer. However, in case of higher niobium concentrations, compositional segregation lead to formation of inhomogeneous passive film and consequently can result in its higher susceptibility to damage by chloride ions [29]. Also in case of CoCrFeNiB alloy, it can be stated that pronounced segregation of chromium contributed to formation of less protective passivation layer.

To further describe the passivation layer formed on the CoCrFeNiSi and CoCrFeNiMo alloys surface, additional measurements were performed in 5% NaCl solution. The obtained Nyquist plots, presented in the **Fig. 11a** and **Fig. 11b**, show large discrepancies. In this case, difference between the CoCrFeNiSi and CoCrFeNiMo is even more pronounced, which is also confirmed by the Bode plots (**Fig. 11c,d**). For the CoCrFeNiSi, a phase angle value in the Bode plot are even closer to the -90°, indicating a highly capacitive response of the specimens. Conversely, for CoCrFeNiMo, the absolute phase angle value decreased, suggesting formation of less stable passivation layer. The spectra obtained for both alloys were fitted using the electric equivalent circuit with two time constants (**Fig. 12**), comprised of solution resistance ($R_s$), passive film resistance ($R_1$), charge transfer resistance ($R_2$), constant phase elements representing the passive film ($CPE_1$) and electrical double-layer capacitance ($CPE_2$) [54]. The values of the fitted electrochemical parameters for impedance data were summarized in **Table 7**. Comparing to 3.5% NaCl environment, the overall polarization resistance of the CoCrFeNiSi passive film noticeably increased, which is in accordance with the results of potentiodynamic polarization measurements. Conversely, in case of the CoCrFeNiMo alloy, the polarization resistance was sustained at almost similar level. However, significant increase of electrical double layer capacitance can be noticed, evidencing increase in surface charge density, facilitating migration of ions across the double layer [54].

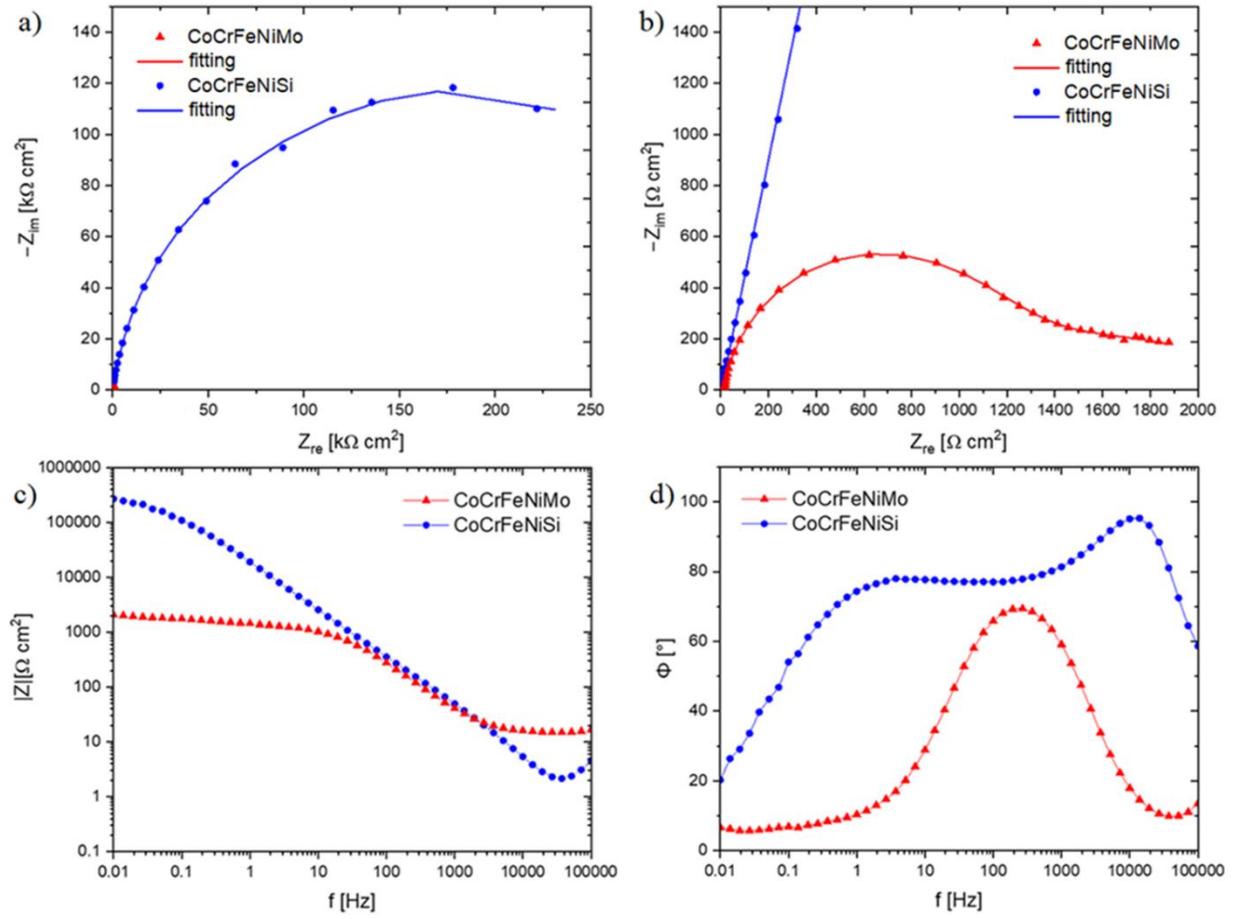

**Fig. 11.** Experimental EIS spectra: Nyquist plots (a) with enlarged initial section of diagram (b), Bode modulus plots (c), Bode phase angle plots (d) for CoCrFeNiMo and CoCrFeNiSi ingots in 5% NaCl solution, at 25 °C.

**Table 7.** The fitted electrochemical parameters for impedance data of the CoCrFeNiMo and CoCrFeNiSi alloys in the as-cast state in 5% NaCl solution, at 25°C.

| Sample | $R_s$ [$\Omega cm^2$] | $CPE_1$ [$\mu\Omega^{-1}cm^{-2}s^n$] | $n_1$ | $R_1$ [$\Omega cm^2$] | $CPE_2$ [$\mu\Omega^{-1}cm^{-2}s^n$] | $n_2$ | $R_1$ [$k\Omega cm^2$] |
|---|---|---|---|---|---|---|---|
| CoCrFeNiMo | 15.26 | 9.44 | 0.91 | 1.02 | 806.8 | 0.35 | 1.32 |
| CoCrFeNiSi | 1.82 | 20.9 | 0.83 | 271.9 | 20.7 | 0.93 | 68.3 |

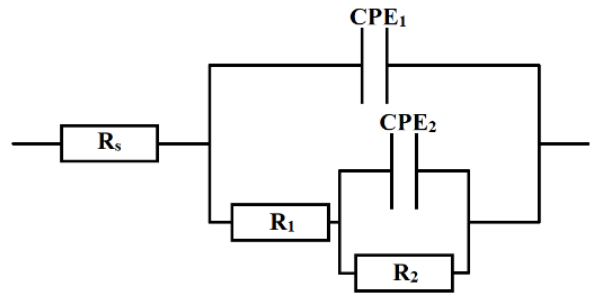

**Fig.12.** Proposed equivalent electric circuit for CoCrFeNiMo and CoCrFeNiSi alloys in 5% NaCl solution at 25 °C

In order to determine the influence of the structure on corrosion resistance, scanning Klevin probe force microscopy studies were carried out to record the differences in the Volta potential on surfaces with ultra-thin water layer. The maps of the Volta potential differences are presented in **Fig. 13**. For the alloy with niobium (**Fig. 13a**), a small difference in potentials can be distinguished between the dendritic $(FeCoCr)_2Nb$ and the interdendritic FCC phases. In the case of the alloy with molybdenum (**Fig. 13b**), the Volta potential difference between the FCC phase and Cr-Mo-TM about 100 mV can be observed. Similarly, the alloy with boron (**Fig. 13c**) was characterized by a clear contrast between $TM_2B$, (Co,Cr)B and FCC matrix. In the case of the CoCrFeNiSi alloy (**Fig. 13d**), for which a binary-phase structure was identified by XRD, while a homogeneous structure was visible in SEM image, local differences in Volta potentials can be observed. These differences may indicate microsegregation that is also visible in the 2D EDX maps (**Fig. 1d**).

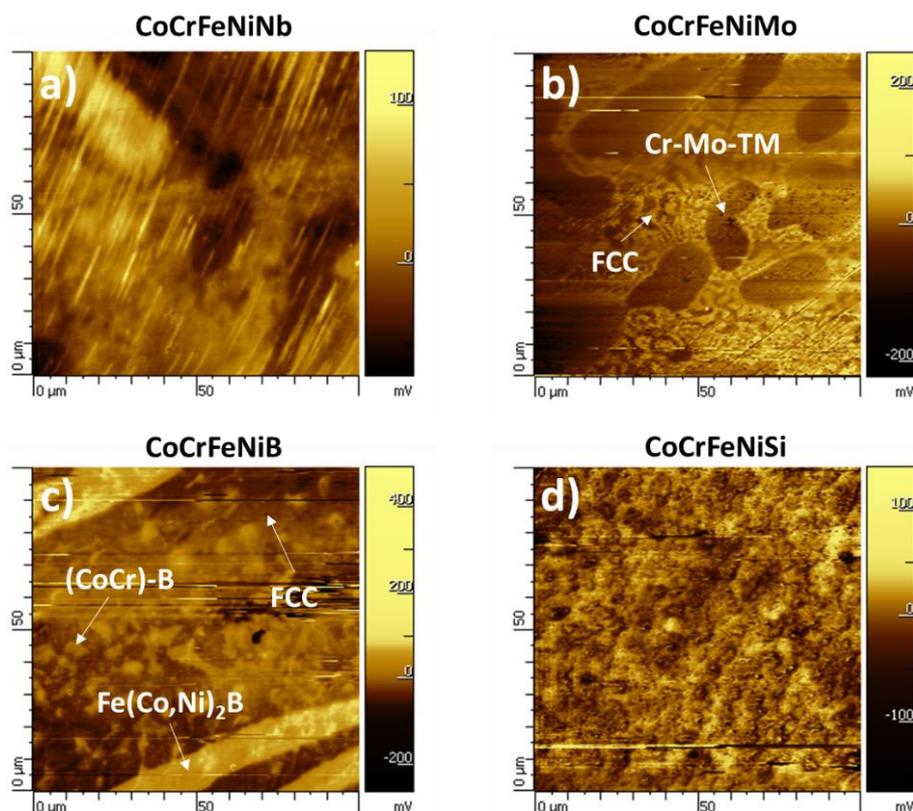

**Fig.13.** SKPFM potential difference and topography maps for CoCrFeNiX (X=Mo,Nb,Si,B) alloys in a form of ingots

Zhang et al. [41] presented the results of the maps of the differences in the Volta potential for alloys of medium and high entropy. The authors [41] indicated that the higher potential of Volta for a given phase means greater nobility. Therefore, the phase with less potential is an anode in galvanic coupling and is susceptible to local corrosion. For the $CoNiVAl_x$ (x = 0, 0.1, 0.2 and 0.3) medium entropy alloys, it was indicated that the FCC phase was anode, and the B2 phase (rich in aluminium and depleted in Ni and V) was cathode. In the case of the FeCoNiCrMo$_{0.3}$ alloy, it was found that the σ phase, rich in Cr, showed greater Volta potential (more noble) than the FCC phase. Linder et al. [57] studied thin films of (CoCrFeNi)1-xMox sputtered by magnetron on a stainless steel substrate. For films containing 0-2 at.% of Mo, the structure consisted of FCC and σ phases, and for higher Mo contents, a single-phase structure was identified. The authors [57] found based on SKPFM maps that there was a large potential difference between FCC and σ phases for films containing 0-2 at.% of Mo. However, due to the small amount of molybdenum, it should be assumed that the σ phase in the films was richer in chromium. Hu et al. [45] described the influence of molybdenum in CoCrFeNiMo$_x$ (x = 0, 0.2, 0.4, 0.6, 0.8, and 1.0) for microgalvanic activity. The authors stated that a higher addition of molybdenum contributes to increasing the differences in Volta potential, where the anode is the FCC phase, and the cathode is the σ phase. In case of this work, it can be assumed that the FCC matrix was characterized by nobler potential, because of nickel enrichment. However, considering the results of electrochemical tests in a solution of NaCl described in this work, we characterized the alloy with molybdenum by the least corrosion resistance, which results from the galvanic corrosion mechanism. A similar corrosion mechanism probably occurred in CoCrFeNiB, for which large differences in the Volta potential were also observed.

Mechanical properties such as hardness and Young's modulus are presented in **Fig. 14a** and **Fig. 14b,** respectively. CoCrFeNiSi and CoCrFeNiB are characterized by the highest hardness among the tested alloys. In the case of the CoCrFeNiB alloy, a relatively high measurement error occurred, due to the presence of borides, which were clearly visible in the SEM images in **Fig. 6.** Furthermore, these precipitates are responsible for an increase in hardness and Young's modulus. In turn, the CoCrFeNiSi alloy is characterised by a uniform structure, and hence, it has little measurement error. The alloys with the addition of Mo and Nb exhibit mechanical parameters that are very similar, both hardness and Young's modulus, but are lower than those of alloys with Si and B. Similarly to the CoCrFeNiB alloy, the presence of many phases influences the hardness and Young's modulus values and contributes to relatively high measurement errors.

According to the literature, molybdenum strengthens the CoCrFeNi alloy through solid solution strengthening and precipitation strengthening mechanisms. The contribution of the secondary phases to the molybdenum alloyed strengthening of the CoCrFeNiMo was investigated in the work [58]. It was found that the Cr-Fe-Mo phase and FCC solid solution showed hardness values of 9.34 GPa and 3.36 GPa, respectively. It confirms the important role of the secondary phase for strengthening of the alloy. Similarly effect was observed in case of CoCrFeNi alloyed with niobium. Chung et al. [59] investigated the effect of niobium on the mechanical properties of CoCrFeNiNb$_{0.5}$, obtaining the average nanohardness value of 9 GPa. The nanohardness of individual phases was estimated to be 13 GPa and 5.5 GPa for (TM)$_2$Nb and FCC solid solution, respectively.

The beneficial effect of Si addition on the strength properties of CoCrFeNi was also shown in the literature. In the work [30], the seven-fold increase in microhardness of the CoCrFeNiSi alloy was obtained, comparing to the alloy without Si addition. It is assumed, that solid solution strengthening, precipitation of the Ni$_x$Si$_y$ intermetallic phase and occurrence of the phase transformation of FCC to BCC phase, synergistically contributed to the alloy strengthening. The phase transformation was related to the decrease in the valence electron concentration (VEC) parameter and the increase in the atomic size difference. The nanomechanical properties of the CoCrFeNiB$_x$ (where x = 0, 0.01, 0.05, 0.1, 0.5) were described in the publication [47]. The most advantageous strength properties were obtained in the case of the CoCrFeNiB$_{0.3}$ alloy, with a nanohardness value of 7.26 GPa. Simultaneously, further increasing the concentration of boron resulted in a decrease of strength properties, which was attributed to reduced structural stability. However, in case of this work, the equimolar CoCrFeNiB characterizes with the significantly higher nanohardness, which probably results from the precipitation strengthening effect caused by formation of borides.

In **Fig. 15a**, load-displacement curves obtained during 30 cycles in a wide load range from 1-500 mN are presented. To determine the elastic-plastic zones in the investigated materials, the stress versus strain curves (presented in **Fig. 15b**) were additionally determined using the 'Filed and Swain" method. This method involves performing multiple load and unload cycles with increasing maximum loads at one point, using a spherical diamond indenter (in our case with radius of R = 25 μm). A series of indentations with a spherical indenter with a gradually higher maximum load gives us results from purely elastic to elastic-plastic. The results presented in **Fig. 15b** show that the alloy with the addition of Nb is the most susceptible to deformation, while the presence of B and Mo strengthens the material and increases its strength, which is important when selecting a material for specific applications in industry.

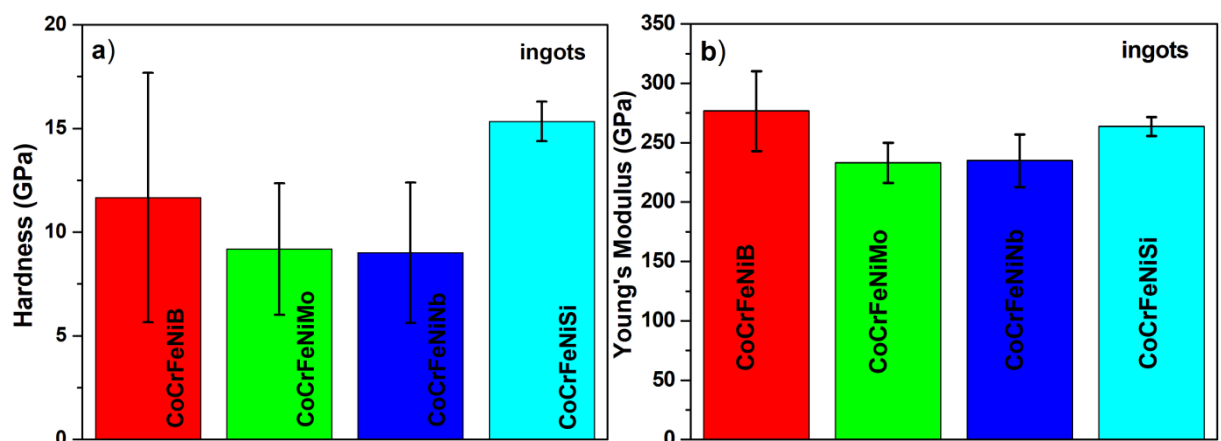

**Fig.14.** Indentation hardness (a) and calculated Young's modulus (b) of CoCrFeNiX (X=Mo,Nb,Si,B) HEAs in a form of ingots

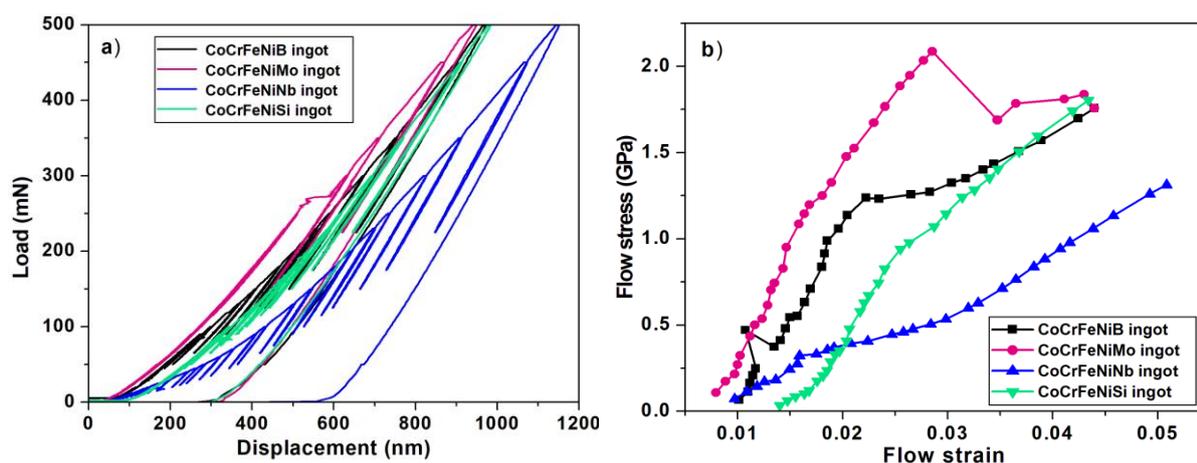

**Fig.15.** Load-displacement curves obtained during 30 cycles of load range from 1-500 mN (a), stress-strain curves extracted from L-D curves of CoCrFeNiX (X=Mo,Nb,Si,B) HEAs in a form of ingots (b)

## 4. Conclusions

The XRD results confirmed that three of the alloys tested contain FCC solid solution, whereas in the case of the CoCrFeNiSi alloy, the formation of two different intermetallic phases was observed. CoCrFeNiNb exhibits a dual phase microstructure, involving an FCC solid solution and a hexagonal (TM)$_2$Nb phase. In the case of the CoCrFeNiMo alloy, two distinct types of Cr-Mo-TM phases were identified along with the solid FCC solution. The most complex phase structure was revealed in the case of B addition. The presence of four different phases was confirmed, including a FCC solid solution and three types of borides. The appearance of Cr-rich and (Ni,Co)-depleted regions can be assigned to the formation of the Cr$_2$B phase, while the orthorhombic Fe(Co,Ni)$_2$B phase was identified in Cr-depleted areas. The presence of a tetragonal (Co,Cr)B phase was also confirmed.

The corrosion behaviour of the alloys was studied in 3.5 and 5% NaCl solutions. The highest corrosion resistance in both solutions used characterizes the CoCrFeNiSi alloy. It can be attributed to its relatively homogeneous elemental distribution, allowing the formation of a stable and protective passive film, which was confirmed by the results of electrochemical impedance spectroscopy measurements. However, still local differences can be observed in the SKPFM maps, although their effect on overall corrosion resistance is limited. In the case of the CoCrFeNiMo alloy, its diminished corrosion resistance can be attributed to the occurrence of pronounced segregation related to the formation of Cr-Mo-TM phases. The resulting potential difference between the Ni-enriched FCC matrix and dendrites constituted of Cr-Mo-TM phases contributed to the occurrence of galvanic coupling corrosion. Although the CoCrFeNiNb alloy also characterizes with substantial heterogeneity in the distribution of elements, especially Cr and Nb, it shows a more favourable corrosion behaviour, exhibiting a wide passivation range during measurements in 5% NaCl solution. The EIS results indicate also the high stability and compactness of the passive film formed. Furthermore, as it can be seen in the SKPFM maps, in this case only slight difference in potentials between the $(TM)_2Nb$ phase and the FCC solid solution can be distinguished. In contrast, the formation of borides in CoCrFeNiB caused substantial segregation, which resulted in a higher susceptibility of the alloy to galvanic coupling corrosion. The uneven distribution of the passivating elements also contributed to the formation of a porous passivation layer with limited protective ability.

The highest mechanical properties were obtained for alloys with the addition of metalloid elements. CoCrFeNiSi is characterized with the highest nanohardness value – exceeding the 15 GPa, while for CoCrFeNiB the highest value of Young modulus value was obtained (above 275 GPa). The CoCrFeNiMo and CoCrFeNiNb alloys exhibit a similar nanohardness and Young modulus.

**Acknowledgements**

The work was supported by the National Science Centre of Poland under research project no. 2022/47/B/ST8/02465.

**Conflict of interest:** All the authors of the article declare that during the implementation of the research presented in the article there was no conflict of interest.

**Ethical approval:** This article does not contain any studies with human participants or animals performed by any of the authors.

**Data availability:** Data will be made available on request.